\def\year{2019}\relax
\documentclass[letterpaper]{article}
\usepackage{aaai20}
\usepackage{times}
\usepackage{helvet}
\usepackage{courier}
\usepackage{url}
\usepackage{graphicx}
\usepackage{multicol}
\usepackage{tabularx}
\usepackage{xcolor}
\usepackage{times}
\usepackage{helvet}
\usepackage{courier}
\usepackage{float}
\usepackage{multirow}
\usepackage{flushend}
\title{A multi-layer approach to disinformation detection on Twitter}
\author{
Francesco Pierri,\textsuperscript{\rm 1}
Carlo Piccardi,\textsuperscript{\rm 1}
Stefano Ceri,\textsuperscript{\rm 1}\\
\textsuperscript{\rm 1}Dipartimento di Elettronica, Informazione e Bioingegneria\\
Politecnico di Milano\\
Via Giuseppe Ponzio, 34, I-20133 Milano, Italy\\
\{firstname.lastname\}@polimi.it
}
\begin{document}

\maketitle
\begin{abstract}
We tackle the problem of classifying news articles pertaining to \textit{disinformation} vs \textit{mainstream} news by solely inspecting their diffusion mechanisms on Twitter. Our technique is inherently simple compared to existing text-based approaches, as it allows to by-pass the multiple levels of complexity which are found in news content (e.g. grammar, syntax, style). We employ a multi-layer representation of Twitter diffusion networks, and we compute for each layer a set of global network features which quantify different aspects of the sharing process. Experimental results with two large-scale datasets, corresponding to diffusion cascades of news shared respectively in the United States and Italy, show that a simple Logistic Regression model is able to classify disinformation vs mainstream networks with high accuracy (AUROC up to 94\%), also when considering the political bias of different sources in the classification task. We also highlight differences in the sharing patterns of the two news domains which appear to be country-independent. We believe that our network-based approach provides useful insights which pave the way to the future development of a system to detect misleading and harmful information spreading on social media.
\end{abstract}

\section{Introduction and related work}
In recent years there has been increasing interest on the issue of disinformation spreading on online social media. Global concern over false (or "fake") news as a threat to modern democracies has been frequently raised--ever since 2016 US Presidential elections--in correspondence of events of political relevance, where the proliferation of manipulated and low-credibility content attempts to drive and influence people opinions \cite{allcott2017}\cite{Grinberg}\cite{Bovet2019}\cite{Lazer18}.

Researchers have highlighted several drivers for the diffusion of such malicious phenomenon, which include human factors (confirmation bias \cite{confirmationbias}, naive realism \cite{naiverealism}), algorithmic biases (\textit{filter bubble} effect \cite{allcott2017}), the presence of deceptive agents on social platforms (bots and trolls \cite{botnature}) and, lastly, the formation of \textit{echo chambers} \cite{spreading2016} where people polarize their opinions as they are insulated from contrary perspectives.

The problem of automatically detecting online disinformation news has been typically formulated as a binary classification task (i.e. credible vs non-credible articles), and tackled with a variety of different techniques, based on traditional machine learning and/or deep learning, which mainly differ in the dataset and the features they employ to perform the classification. We may distinguish three approaches: those built on content-based features, those based on features extracted from the social context, and those which combine both aspects. A few main challenges hinder the task, namely the impossibility to manually verify all news items, the lack of gold-standard datasets and the adversarial setting in which malicious content is created \cite{Lazer18}\cite{botnature}.

In this work we follow the direction pointed out in a few recent contributions on the diffusion of disinformation compared to traditional and objective information. These have shown that false news spread faster and deeper than true news \cite{Vosoughi18}, and that social bots and echo chambers play an important role in the diffusion of malicious content \cite{botnature,spreading2016}. Therefore we focus on the analysis of spreading patterns which naturally arise on social platforms as a consequence of multiple interactions between users, due to the increasing trend in online sharing of news \cite{allcott2017}.

A deep learning framework for detection of fake news cascades is provided in \cite{monti2019fake}, where the authors refer to \cite{Vosoughi18} in order to collect Twitter cascades pertaining to verified false and true rumors. They employ \textit{geometric} deep learning, a novel paradigm for graph-based structures, to classify cascades based on four categories of features, such as user profile, user activity, network and spreading, and content. They also observe that a few hours of propagation are sufficient to distinguish false news from true news with high accuracy. Diffusion cascades on Weibo and Twitter are analyzed in \cite{chinafake}, where authors focus on highlighting different topological properties, such as the number of \textit{hops} from the source or the heterogeneity of the network, to show that fake news shape diffusion networks which are highly different from credible news, even at early stages of propagation.

In this work, we consider the results of \cite{PierriScirep2019} as our baseline. The authors use off-the-shelf machine learning classifiers to accurately classify news articles leveraging Twitter diffusion networks. To this aim, they consider a set of basic features which can be qualitatively interpreted w.r.t to the social behavior of users sharing credible vs non-credible information. Their methodology is overall in accordance with \cite{astroturfing}, where authors successfully detect Twitter \textit{astroturfing} content, i.e. political campaigns disguised as spontaneous grassroots, with a machine learning framework based on network features.

In this paper, we propose a classification framework based on a multi-layer formulation of Twitter diffusion networks.
For each article we disentangle different social interactions on Twitter, namely tweets, retweets, mentions, replies and quotes, to accordingly build a diffusion network composed of multiple layers (on for each type of interaction), and we compute structural features separately for each layer. We pick a set of global network properties from the network science toolbox which can be qualitatively explained in terms of social dimensions and allow us to encode different networks with a tuple of features. These include traditional indicators, e.g. network density, number of strong/weak connected components and diameter, and more elaborated ones such as main K-core number \cite{k-core} and structural virality \cite{goel}. Our main research question is whether the use of a multi-layer, disentangled network yields a significant advance in terms of classification accuracy over a conventional single-layer diffusion network. Additionally, we are interested in understanding which of the above features, and in which layer, are most effective in the classification task.

We perform classification experiments with an off-the-shelf Logistic Regression model on two different datasets of mainstream and disinformation news shared on Twitter respectively in the United States and in Italy during 2019.
In the former case we also account for political biases inherent to different news sources, referring to the procedure proposed in \cite{Bovet2019} to label different outlets. Overall we show that we are able to classify credible vs non-credible diffusion networks (and consequently news articles) with high accuracy (AUROC up to 94\%), even when accounting for the political bias of sources (and training only on left-biased or right-biased articles). We observe that the layer of mentions alone conveys useful information for the classification, denoting a different usage of this functionality when sharing news belonging to the two news domains. We also show that most discriminative features, which are relative to the breadth and depth of largest cascades in different layers, are the same across the two countries.

The outline of this paper is the following: we first formulate the problem and describe data collection, network representation and structural properties employed for the classification; then we provide experimental results--classification performances, layer and feature importance analyses and a temporal classification evaluation--and finally we draw conclusions and future directions.

\section{Methodology}
\subsection{Disinformation and mainstream news}
In this work we formulate our classification problem as follows: given two classes of news articles, respectively $D$ (\textit{disinformation}) and $M$ (\textit{mainstream}), a set of news articles $A_i$ and associated class labels $C_i \in \{D,M\}$, and a set of tweets $\Pi_i=\{T_i^1, T_i^2, ...\}$ each of which contains an Uniform Resource Locator (URL) pointing explicitly to article $A_i$, predict the class $C_i$ of each article $A_i$. 

There is huge debate and controversy on a proper taxonomy of malicious and deceptive information \cite{Grinberg}\cite{Bovet2019}\cite{botometer}\cite{hoaxy}\cite{misinformation}\cite{Lazer18}\cite{PierriScirep2019}. In this work we prefer the term \textit{disinformation} to the more specific \textit{fake news} to refer to a variety of misleading and harmful information. Therefore, we follow a \textit{source-based} approach, a consolidated strategy also adopted by \cite{botnature}\cite{hoaxy}\cite{Bovet2019}\cite{Grinberg}, in order to obtain relevant data for our analysis. We collected:
\begin{enumerate}
\item Disinformation articles, published by websites which are well-known for producing low-credibility content, false and misleading news reports as well as extreme propaganda and hoaxes and flagged as such by reputable journalists and fact-checkers;
\item Mainstream news, referring to traditional news outlets which deliver factual and credible information.
\end{enumerate}
We believe that this is currently the most reliable classification approach, but it entails obvious limitations, as disinformation outlets may also publish true stories and likewise misinformation is sometimes reported on mainstream media. Also, given the choice of news sources, we cannot test whether our methodology is able to classify disinformation vs factual but not mainstream news which are published on niche, non-disinformation outlets.

\begin{figure}[!t]
    \centering
    \includegraphics[width=\linewidth]{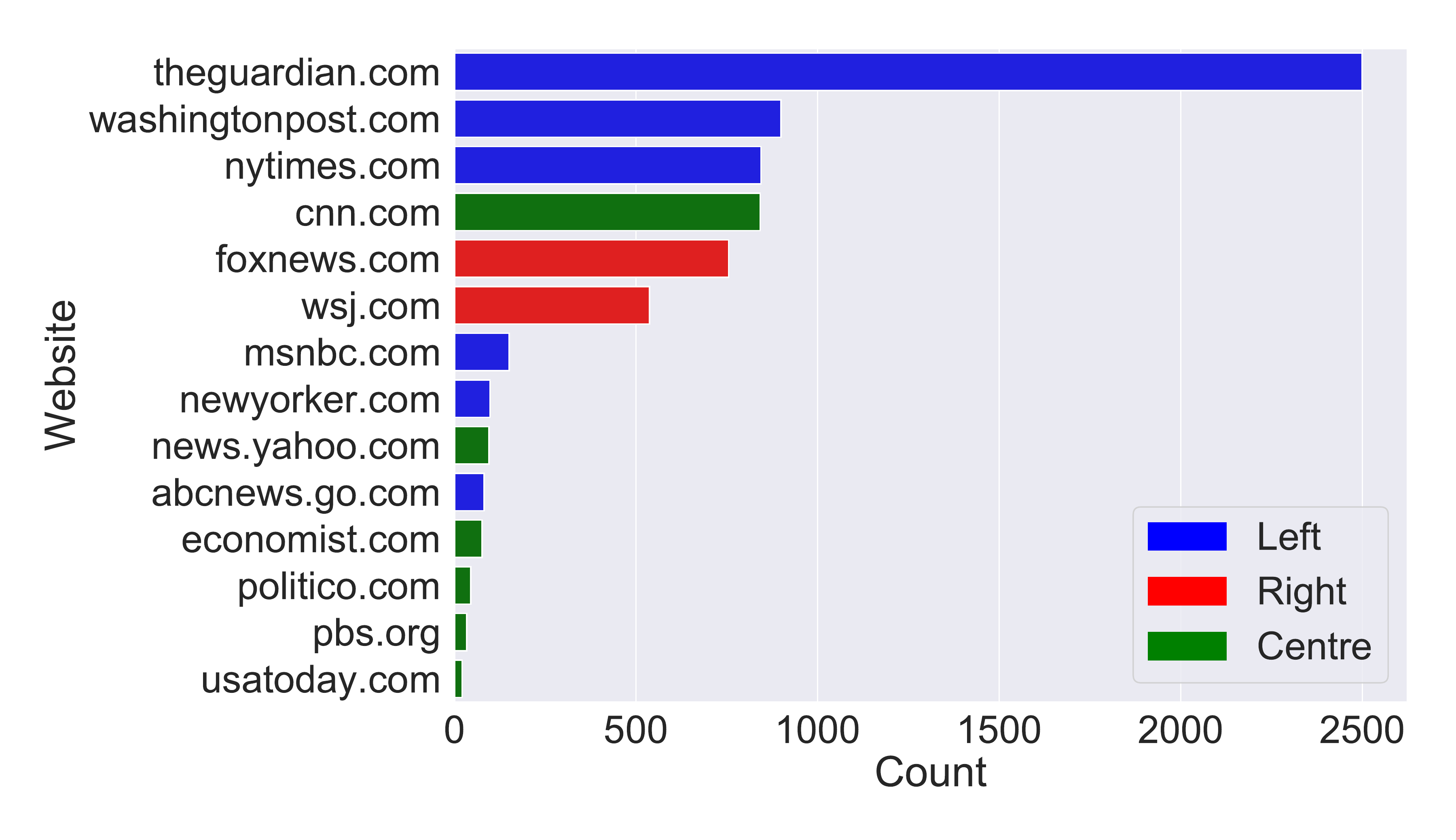}
    \\
    \footnotesize{\textbf{a)} US mainstream articles}
    \\
    \includegraphics[width=\linewidth]{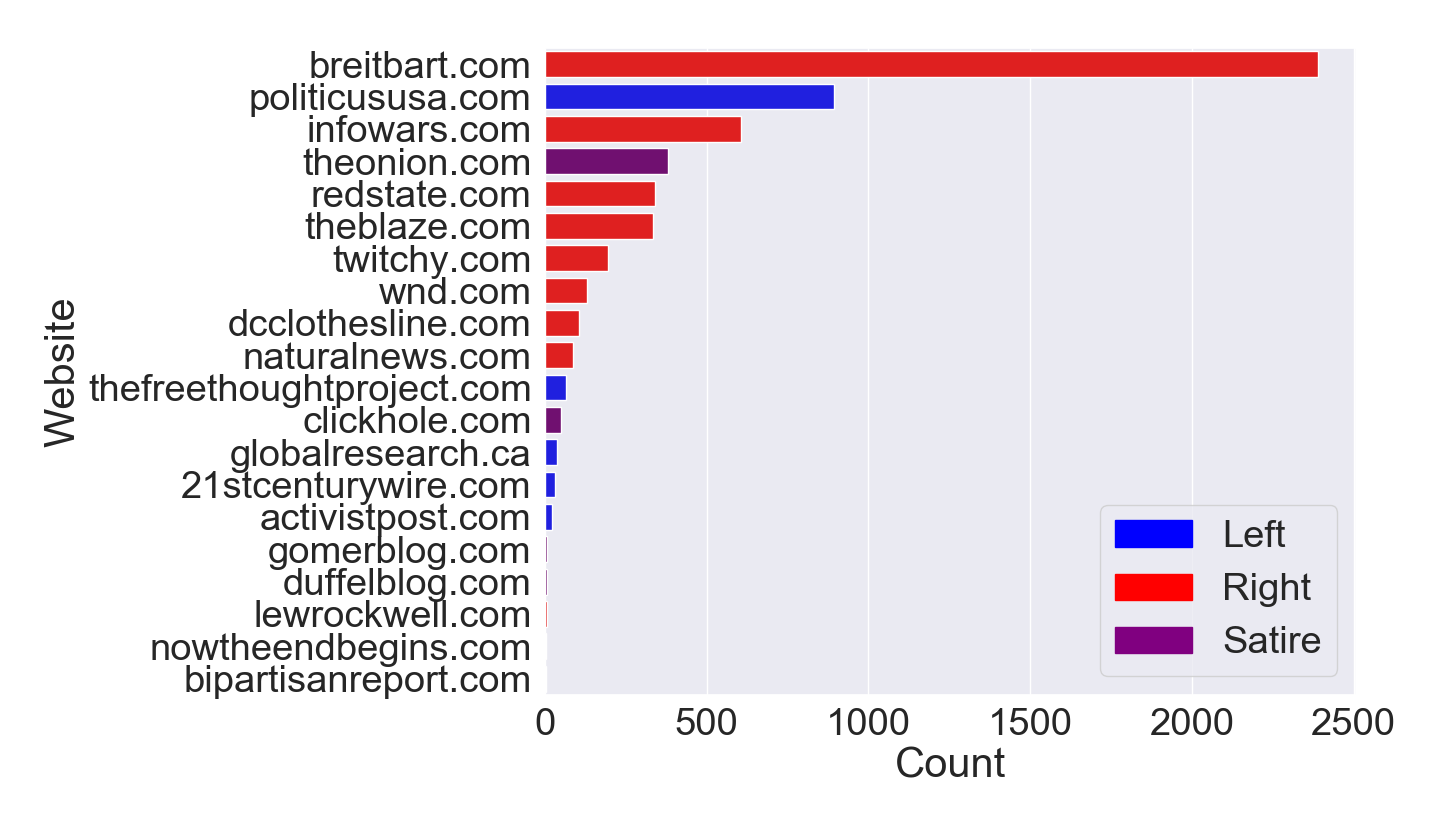}
    \\
    \footnotesize{\textbf{b)} US disinformation articles}
    \caption{Distribution of the number of articles per source for US \textbf{a)} mainstream and \textbf{b)} disinformation news. Colors indicate the political bias label of each source.}
    \label{fig:breakdown}
\end{figure}

\begin{figure}[!t]
    \centering
    \includegraphics[width=\linewidth]{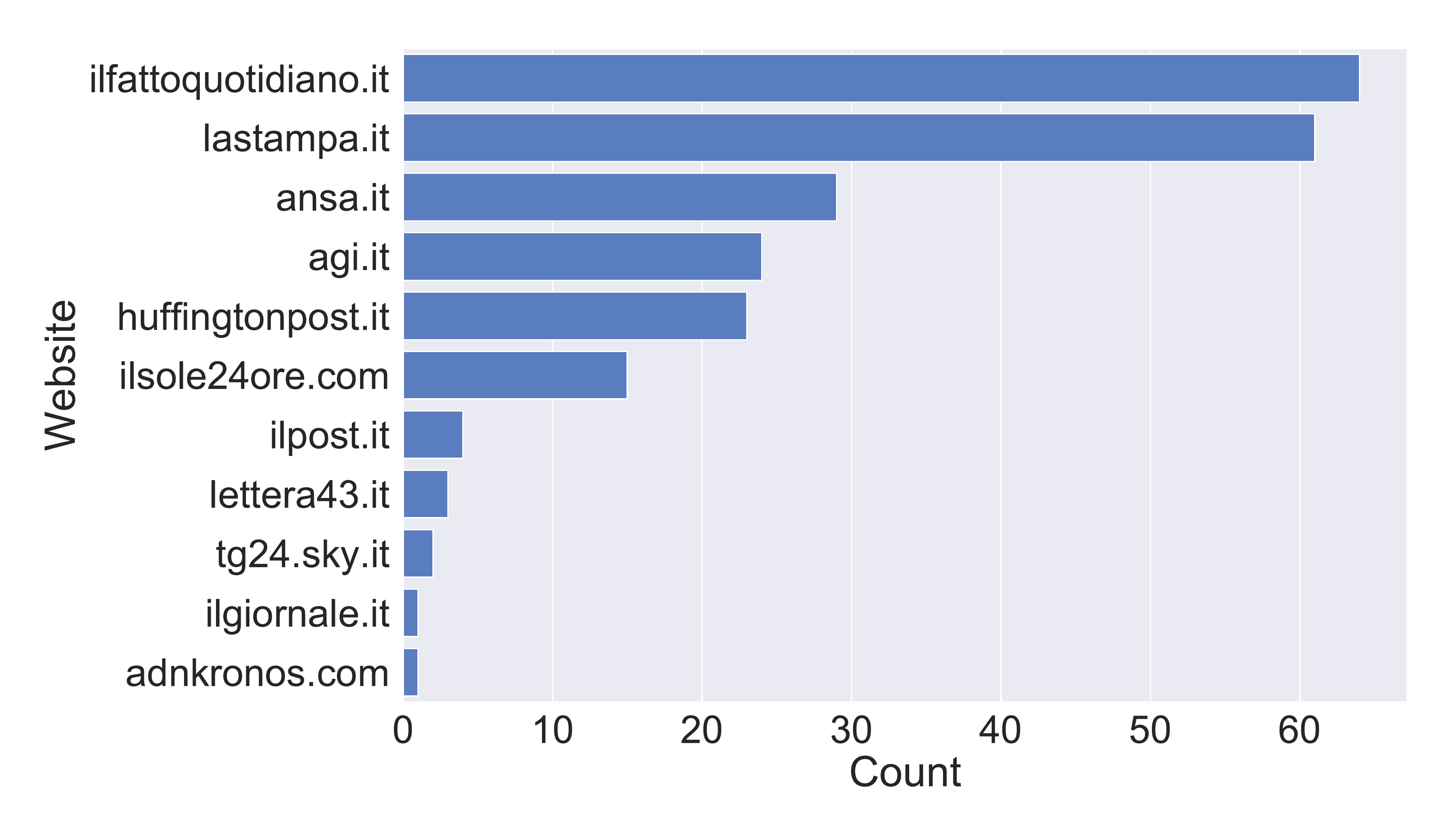}
    \\
    \footnotesize{\textbf{a)} IT mainstream articles}
    \\
    \includegraphics[width=\linewidth]{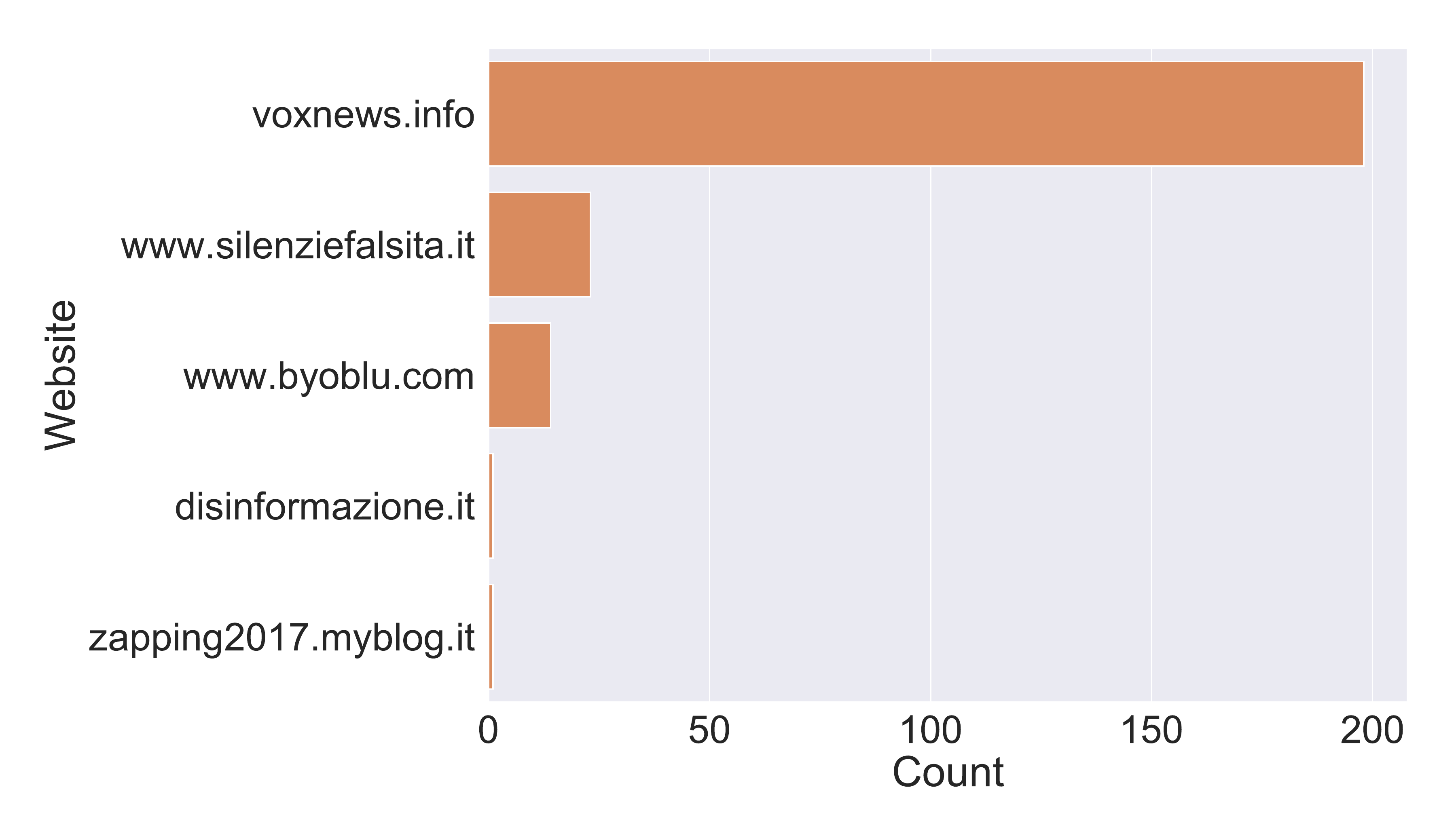}
    \\
    \footnotesize{\textbf{b)} IT disinformation articles}
    \caption{Distribution of the number of articles per source for Italian \textbf{a)} mainstream and \textbf{ b)} disinformation news.}
    \label{fig:breakdown-ita}
\end{figure}

\subsection{US dataset}
We collected tweets associated to a dozen US \textit{mainstream} news websites, i.e. most trusted sources described in \cite{pew}, with the Streaming API, and we referred to Hoaxy API \cite{hoaxy} for what concerns tweets containing links to 100+ US \textit{disinformation} outlets. We filtered out articles associated to less than 50 tweets. The resulting dataset contains overall $\sim$1.7 million tweets for mainstream news, collected in a period of three weeks (February 25th, 2019-March 18th, 2019), which are associated to 6,978 news articles, and $\sim$1.6 million tweets for disinformation, collected in a period of three months (January 1st, 2019-March 18th, 2019) for sake of balance of the two classes, which hold 5,775 distinct articles. Diffusion censoring effects \cite{goel} were correctly taken into account in both collection procedures.
We provide in Figure \ref{fig:breakdown} the distribution of articles by source and political bias for both news domains.

As it is reported that conservatives and liberals exhibit different behaviors on online social platforms \cite{barbera}\cite{conover2012}\cite{bovet2018bis}, we further assigned a political bias label to different US outlets (and therefore news articles) following the procedure described in \cite{Bovet2019}. In order to assess the robustness of our method, we performed classification experiments by training only on left-biased (or right-biased) outlets of both disinformation and mainstream domains and testing on the entire set of sources, as well as excluding particular sources that outweigh the others in terms of samples to avoid over-fitting.

\subsection{Italian dataset}
For what concerns the Italian scenario we first collected tweets with the Streaming API in a 3-week period (April 19th, 2019-May 5th, 2019), filtering those containing URLs pointing to Italian official newspapers websites as described in \cite{vicario2017}; these correspond to the list provided by the association for the verification of newspaper circulation in Italy (Accertamenti Diffusione Stampa)\footnote{\url{http://www.adsnotizie.it}}.
We instead referred to the dataset provided by \cite{PierriArtoni2019} to obtain a set of tweets, collected continuously since January 2019 using the same Twitter endpoint, which contain URLs to 60+ Italian disinformation websites\footnote{The list is available at \url{https://bit.ly/30lJKhx}}. In order to get balanced classes (April 5th, 2019-May 5th, 2019), we retained data collected in a longer period w.r.t to mainstream news. In both cases we filtered out articles with less than 50 tweets; overall this dataset contains $\sim$160k mainstream tweets, corresponding to 227 news articles, and $\sim$100k disinformation tweets, corresponding to 237 news articles. We provide in Figure \ref{fig:breakdown-ita} the distribution of articles according to distinct sources for both news domains.
As in the US dataset, we took into account censoring effects \cite{goel} by excluding tweets published before (\textit{left-censoring}) or after two weeks (\textit{right-censoring}) from the beginning of the collection process.

The different volumes of news shared on Twitter in the two countries are due both to the different population size of US and Italy (320 vs 60 millions) but also to the different usage of Twitter platform (and social media in general) for news consumption \cite{reuters2019}. Both datasets analyzed in this work are available from the authors on request.

A crucial aspect in our approach is the capability to fully capturing sharing cascades on Twitter associated to news articles. It has been reported \cite{twittersample} that the Twitter streaming endpoint filters out tweets matching a given query if they exceed 1\% of the global daily volume\footnote{\url{https://www.internetlivestats.com/twitter-statistics/}} of shared tweets, which nowadays is approximately $5\cdot10^8$; however, as we always collected less than $10^6$ tweets per day, we did not incur in this issue and we thus gathered 100\% of tweets matching our query.

\begin{figure}[!t]
    \centering
    \includegraphics[width=\linewidth]{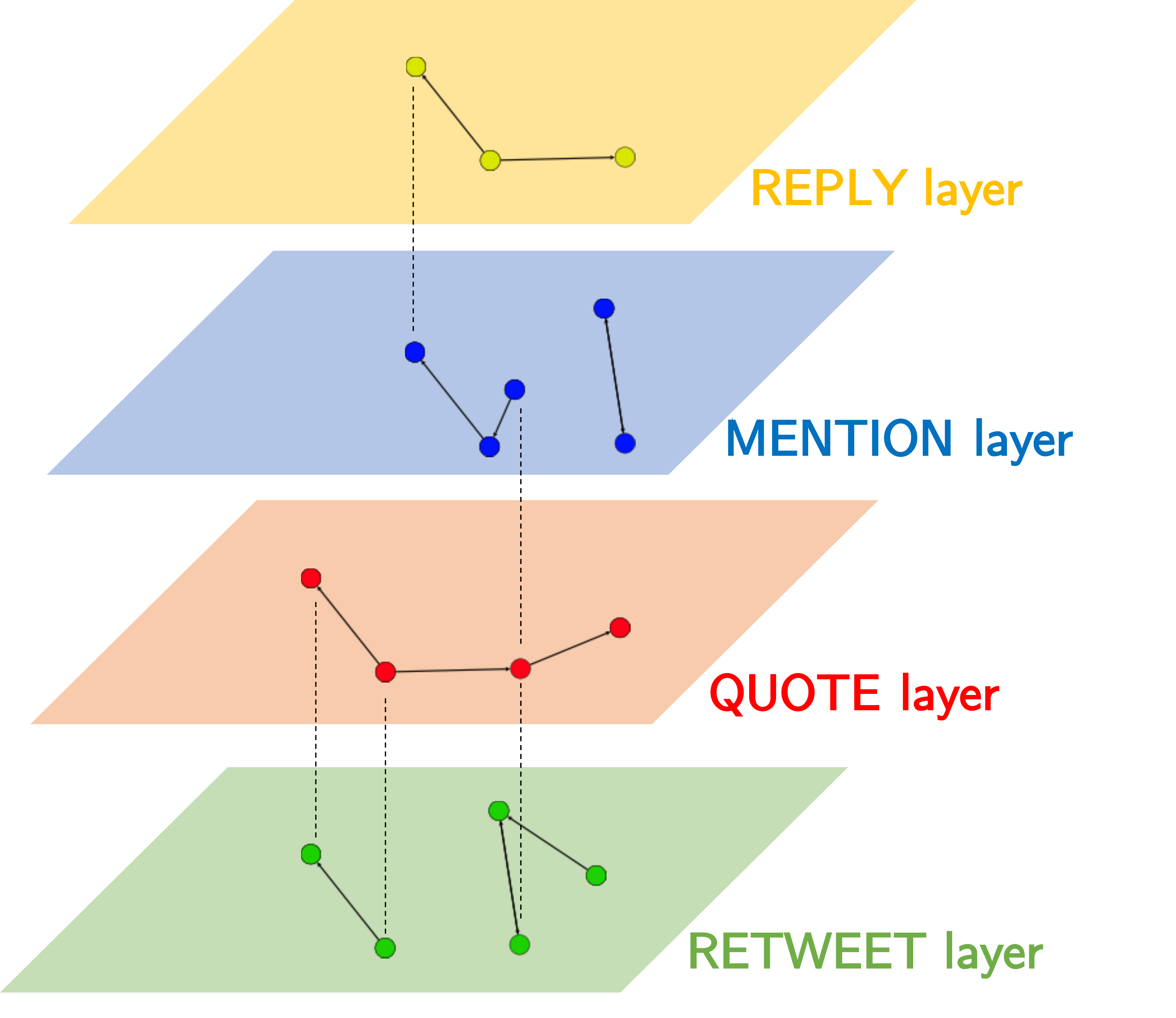}
    \caption{An example of Twitter multi-layer diffusion network with four layers.}
    \label{fig:multi}
\end{figure}

\subsection{Building diffusion networks}
We built Twitter diffusion networks following an approach widely adopted in the literature \cite{botnature}\cite{misinformation}\cite{Bovet2019}. We remark that there is an unavoidable limitation in Twitter Streaming API, which does not allow to retrieve \textit{true} re-tweeting cascades because re-tweets always point to the original source and not to intermediate re-tweeting users \cite{Vosoughi18}\cite{goel}; thus we adopt the only viable approach based on Twitter's public availability of data. Besides, by disentangling different interactions with multiple layers we potentially reduce the impact of this limitation on the global network properties compared to the single-layer approach used in our baseline.

Using the notation described in \cite{multilayer}. we employ a multi-layer representation for Twitter diffusion networks. Sociologists have indeed recognized decades ago that it is crucial to study social systems by constructing multiple social networks where different types of ties among same individuals are used \cite{sociologists}.
Therefore, for each news article we built a multi-layer diffusion network composed of four different layers, one for each type of social interaction on Twitter platform, namely retweet (RT), reply (R), quote (Q) and mention (M), as shown in Figure \ref{fig:multi}. These networks are not necessarily \textit{node-aligned}, i.e. users might be missing in some layers. We do not insert "dummy" nodes to represent all users as it would have severe impact on the global network properties (e.g. number of weakly connected components). 
Alternatively one may look at each multi-layer diffusion network as an ensemble of individual graphs \cite{multilayer}; since global network properties are computed separately for each layer, they are not affected by the presence of any \textit{inter-layer} edges.

In our multi-layer representation, each layer is a directed graph where we add edges and nodes for each tweet of the layer type, e.g. for the RT layer: whenever user $a$ retweets account $b$ we first add nodes $a$ and $b$ if not already present in the RT layer, then we build an edge that goes from $b$ to $a$ if it does not exists or we increment the weight by 1. Similarly for the other layers: for the R layer edges go from user $a$ (who replies) to user $b$, for the Q layer edges go from user $b$ (who is quoted by) to user $a$ and for the M layer edges go from user $a$ (who mentions) to user $b$.

Note that, by construction, our layers do not include isolated nodes; they correspond to "pure tweets", i.e. tweets which have not originated any interactions with other users. However, they are present in our dataset, and their number is exploited for classification, as described below.

\subsection{Global network properties}
We used a set of global network indicators which allow us to encode each network layer by a tuple of features. Then we simply concatenated tuples as to represent each multi-layer network with a single feature vector. We used the following global network properties: \begin{enumerate}
\item \textbf{Number of Strongly Connected Components (SCC)}: a Strongly Connected Component of a directed graph is a maximal (sub)graph where for each pair of vertices $u,v$ there is a path in each direction ($u\rightarrow v$, $v\rightarrow u$).
\item \textbf{Size of the Largest Strongly Connected Component (LSCC)}: the number of nodes in the largest strongly connected component of a given graph.
\item \textbf{Number of Weakly Connected Components (WCC)}: a Weakly Connected Component of a directed graph is a maximal (sub)graph where for each pair of vertices $(u, v)$ there is a path $u \leftrightarrow v$ ignoring edge directions.
\item \textbf{Size of the Largest Weakly Connected Component (LWCC)}: the number of nodes in the largest weakly connected component of a given graph.
\item \textbf{Diameter of the Largest Weakly Connected Component (DWCC)}: the largest distance (length of the shortest path) between two nodes in the (undirected version of) largest weakly connected component of a graph.
\item \textbf{Average Clustering Coefficient (CC)}: the average of the local clustering coefficients of all nodes in a graph; the local clustering coefficient of a node quantifies how close its neighbours are to being a complete graph (or a clique). It is computed according to \cite{cc}.
\item \textbf{Main K-core Number (KC)}: a K-core \cite{k-core} of a graph is a maximal sub-graph that contains nodes of internal degree $k$ or more; the main K-core number is the highest value of $k$ (in directed graphs the total degree is considered).
\item \textbf{Density (d)}: the density for directed graphs is $d=\frac{|E|}{|V||V-1|}$, where $|E|$ is the number of edges and $|N|$ is the number of vertices in the graph; the density equals 0 for a graph without edges and 1 for a complete graph.
\item \textbf{Structural virality of the largest weakly connected component (SV)}: this measure is defined in \cite{goel} as the average distance between all pairs of nodes in a cascade tree or, equivalently, as the average depth of nodes, averaged over all nodes in turn acting as a root; for $|V| > 1$ vertices, $SV=\frac{1}{|V||V-1|}\sum_i\sum_j d_{ij}$ where $d_{ij}$ denotes the length of the shortest path between nodes $i$ and $j$. This is equivalent to compute the Wiener's index \cite{wiener} of the graph and multiply it by a factor $\frac{1}{|V||V-1|}$. In our case we computed it for the undirected equivalent graph of the largest weakly connected component, setting it to 0 whenever $V=1$.
\end{enumerate}
We used \verb|networkx| Python package \cite{networkx} to compute all features. Whenever a layer is empty. we simply set to 0 all its features.
In addition to computing the above nine features for each layer, we added two indicators for encoding information about pure tweets, namely the number \textbf{T} of pure tweets (containing URLs to a given news article) and the number \textbf{U} of unique users authoring those tweets. Therefore, a single diffusion network is represented by a vector with $9\cdot4+2=38$ entries.

\subsection{Interpretation of network features and layers}
Aforementioned network properties can be qualitatively explained in terms of social footprints as follows:
\textbf{SCC} correlates with the size of the diffusion network, as the propagation of news occurs in a broadcast manner most of the time, i.e. re-tweets dominate on other interactions, while \textbf{LSCC} allows to distinguish cases where such mono-directionality is somehow broken.  \textbf{WCC} equals (approximately) the number of distinct diffusion cascades pertaining to each news article, with exceptions corresponding to those cases where some cascades merge together via Twitter interactions such as mentions, quotes and replies, and accordingly \textbf{LWCC} and \textbf{DWCC} equals the size and the depth of the largest cascade. \textbf{CC} corresponds to the level of connectedness of neighboring users in a given diffusion network whereas \textbf{KC} identifies the set of most influential users in a network and describes the efficiency of information spreading \cite{misinformation}. Finally, \textbf{d} describes the proportions of potential connections between users which are actually activated and \textbf{SV} indicates whether a news item has gained popularity with a single and large broadcast or in a more viral fashion through multiple generations.

For what concerns different Twitter actions, users primarily interact with each other using retweets and mentions \cite{conover2012}. \\
The former are the main engagement activity and act as a form of endorsement, allowing users to rebroadcast content generated by other users \cite{boyd2010}. Besides, when node B retweets node A we have an implicit confirmation that information from A appeared in B's Twitter feed \cite{astroturfing}. Quotes are simply a special case of retweets with comments.\\
Mentions usually include personal conversations as they allow someone to address a specific user or to refer to an individual in the third person; in the first case they are located at the beginning of a tweet and they are known as replies, otherwise they are put in the body of a tweet \cite{conover2012}. The network of mentions is usually seen as a stronger version of interactions between Twitter users, compared to the traditional graph of follower/following relationships \cite{mentions}.

\section{Experiments}
\subsection{Setup}
We performed classification experiments using a basic off-the-shelf classifier, namely Logistic Regression (LR) with L2 penalty; this also allows us to compare results with our baseline. We applied a standardization of the features and we used the default configuration for parameters as described in \verb|scikit-learn| package \cite{scikit}. We also tested other classifiers (such as K-Nearest Neighbors, Support Vector Machines and Random Forest) but we omit results as they give comparable performances. We remark that our goal is to show that a very simple machine learning framework, with no parameter tuning and optimization, allows for accurate results with our network-based approach.

\begin{table}[!t]
\centering
\resizebox{\linewidth}{!}{%
\begin{tabular}{l|l|l|l|l|l|l}
 \multicolumn{1}{c}{}& \multicolumn{3}{c|}{\textbf{No. Mainstream}} & \multicolumn{3}{c}{\textbf{No. Disinformation}} \\
\multicolumn{1}{l|}{\textbf{Size Class}} & \multicolumn{1}{l|}{Left} & \multicolumn{1}{l|}{Right} & \multicolumn{1}{l|}{Tot.} & \multicolumn{1}{l|}{Left} & \multicolumn{1}{l|}{Right} & \multicolumn{1}{l}{Tot.} \\ \hline
 $[0, 100)$ & 774 & 2746 & 4177 & 379 & 2086 & 2640 \\
 $[100, 1000)$ & 1712 & 464 & 2605 & 654 & 1946 & 2900\\
 $[1000, + \infty)$ & 115 & 54 & 196 & 19 & 162 & 235 \\
 $[0, +\infty)$ & 4573  & 1292  & 6978  & 1052 & 4194  & 5575 \\
 \hline
\end{tabular}
}
\caption{Composition of the US dataset according to class, size and political bias.} 
\end{table}
\begin{table}[!t]
\centering
\resizebox{\linewidth}{!}{%
\begin{tabular}{l|l|l}
\textbf{Size Class} &  \textbf{No. Mainstream} & \textbf{No. Disinformation} \\ \hline
$[0, 100)$ & 165 & 79 \\
$[100, 1000)$ & 61 & 158 \\
$[0, +\infty)$ & 227 & 237 \\
\hline
\end{tabular}%
}
\caption{Composition of the Italian dataset according to class and size.}
\end{table}


We used the following evaluation metrics to assess the performances of different classifiers (TP=true positives, FP=false positives, FN=false negatives):
\begin{enumerate}
    \item \textbf{Precision} = $\frac{TP}{TP+FP}$, the ability of a classifier not to label as positive a negative sample.
    \item \textbf{Recall} =  $\frac{TP}{TP+FN}$, the ability of a classifier to retrieve all positive samples.
    \item \textbf{F1-score} = $2 \frac{\mbox{Precision} \cdot \mbox{Recall}}{\mbox{Precision} + \mbox{Recall}}$, the harmonic average of Precision and Recall.
    \item \textbf{Area Under the Receiver Operating Characteristic curve (AUROC)}; the Receiver Operating Characteristic (ROC) curve \cite{roc}, which plots the TP rate versus the FP rate, shows the ability of a classifier to discriminate positive samples from negative ones as its threshold is varied; the AUROC value is in the range $[0, 1]$, with the random baseline classifier holding AUROC$=0.5$ and the ideal perfect classifier AUROC$=1$; thus larger AUROC values (and steeper ROCs) correspond to better classifiers.
\end{enumerate}
In particular we computed so-called \textit{macro} average--simple unweighted mean--of these metrics evaluated considering both labels (\textit{disinformation} and \textit{mainstream}). We employed stratified shuffle split cross validation (with 10 folds) to evaluate performances.

Finally,
we partitioned networks according to the total number of unique users involved in the sharing, i.e. the number of nodes in the aggregated network represented with a single-layer representation considering together all layers and also pure tweets. A breakdown of both datasets according to size class (and political biases for the US scenario) is provided in Table 1 and Table 2.

\begin{table}[!t]
\resizebox{\linewidth}{!}{%
\begin{tabular}{l|l|l|l|l}
\textbf{Size Class} & \textbf{AUROC} & \textbf{Precision} & \textbf{Recall} & \textbf{F1-score} \\
\hline
(US) $[0, 100)$ & 0.87 $\pm$ 0.01
 & 0.79 $\pm$ 0.01 & 0.77 $\pm$ 0.01 & 0.78 $\pm$ 0.01
\\ \hline
(US) $[100, 1000)$ &
0.93 $\pm$ 0.01 & 0.87 $\pm$ 0.01 & 0.87 $\pm$ 0.01 & 0.87 $\pm$ 0.01
\\ \hline
(US) $[1000, +\infty)$ & 0.94 $\pm$ 0.02 & 0.86 $\pm$ 0.05 & 0.86 $\pm$ 0.05 & 0.86 $\pm$ 0.05
\\ \hline
(US) $[0, +\infty)$ & 0.88 $\pm$ 0.01 & 0.81 $\pm$ 0.01 & 0.80 $\pm$ 0.01 & 0.80 $\pm$ 0.01
\\ \hline \hline
(IT) $[0, 100)$ &
0.89 $\pm$ 0.06
 &
0.81 $\pm$ 0.11
 &
0.82 $\pm$ 0.11
 &
0.81 $\pm$ 0.11
\\ \hline
(IT) $[100, 1000)$ &
0.86 $\pm$ 0.07
 &
0.83 $\pm$ 0.08
 &
0.78 $\pm$ 0.06
 &
0.80 $\pm$ 0.06
\\ \hline
(IT) $[0, +\infty)$ &
0.90 $\pm$ 0.02
 &
0.81 $\pm$ 0.05
 &
0.81 $\pm$ 0.05
 &
0.81 $\pm$ 0.05
\\ \hline
\end{tabular}
}
\caption{Different evaluation metrics for the LR classifier (using a multi-layer approach) evaluated on different size classes of both the US and the Italian dataset.}
\end{table}

\subsection{Classification performances}
In Table 3 we first provide classification performances on the US dataset for the LR classifier evaluated on the size class described in Table 1. We can observe that in all instances our methodology performs better than a random classifier (50\% AUROC), with AUROC values above 85\% in all cases. 

\begin{table}[!t]
\resizebox{\linewidth}{!}{%
\begin{tabular}{l|l|l}
\textbf{Size Class} & \textbf{Single-layer} & \textbf{Multi-layer} \\
\hline
(US) $[0, 100)$ & 0.74 $\pm$ 0.02 & 0.87 $\pm$ 0.01\\
\hline
(US) $[100, 1000)$ & 0.85 $\pm$ 0.02 & 0.93 $\pm$ 0.01\\
\hline
(US) $[1000, +\infty)$ & 0.93 $\pm$ 0.03 & 0.94 $\pm$ 0.02\\
\hline
(US) $[0, +\infty)$ & 0.78 $\pm$ 0.02 & 0.88 $\pm$ 0.01\\
\hline
\hline
(IT) $[0, 100)$ & 0.77 $\pm$ 0.08 & 0.89 $\pm$ 0.06 \\
\hline
(IT) $[100, 1000)$ & 0.66 $\pm$ 0.14 & 0.86 $\pm$ 0.07 \\
\hline
(IT) $[0, +\infty)$ & 0.74 $\pm$ 0.12 & 0.90 $\pm$ 0.02 \\
\hline
\end{tabular}
}
\caption{Comparison of performances of our multi-layer approach vs the baseline (single-layer). We show AUROC values for the LR classifier evaluated on different size classes of both US and IT datasets.}
\end{table}

\begin{figure}[!t]
    \centering
    \includegraphics[width=\linewidth]{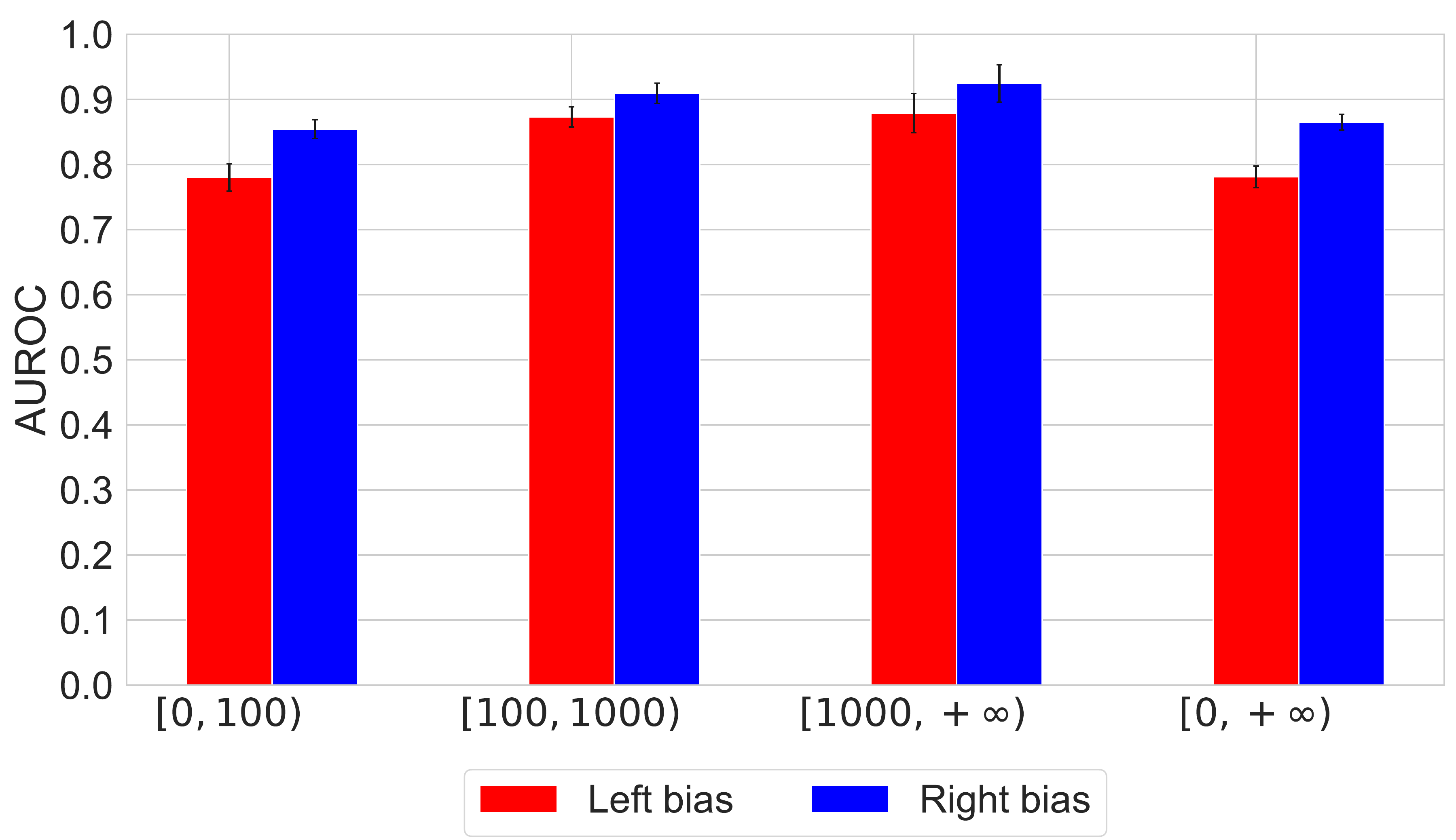}
    \caption{AUROC values for the Balanced Random Forest classifier trained on left-biased (red) and right-biased (blue) news articles in the US dataset, and tested on the entire dataset. Error bars indicate the standard deviation of AUROC values over different folds of the cross validation.}
\end{figure}

For what concerns political biases,
as the classes of mainstream and disinformation networks are not balanced (e.g., 1,292 mainstream and 4,149 disinformation networks with right bias) we employ a Balanced Random Forest with default parameters (as provided in \verb|imblearn| Python package \cite{imbalanced}). In order to test the robustness of our methodology, we trained only on left-biased networks or right-biased networks and tested on the entire set of sources (relative to the US dataset); we provide a comparison of AUROC values for both biases in Figure 4. We can notice that our multi-layer approach still entails significant results, thus showing that it can accurately distinguish mainstream news from disinformation regardless of the political bias. We further corroborated this result with additional classification experiments, that show similar performances, in which we excluded from the training/test set two specific sources (one at a time and both at the same time) that outweigh the others in terms of data samples--respectively "breitbart.com" for right-biased sources and "politicususa.com" for left-biased ones.


We performed classification experiments on the Italian dataset using the LR classifier and different size classes (we excluded $[1000, +\infty)$ which is empty); we show results for different evaluation metrics in Table 3. We can see that despite the limited number of samples (one order of magnitude smaller than the US dataset) the performances are overall in accordance with the US scenario.

As shown in Table 4, we obtain results which are much better than our baseline in all size classes (see Table 4):
\begin{itemize}
    \item In the US dataset our multi-layer methodology performs much better in all size classes except for large networks ($[1000, +\infty)$ size class), reaching up to 13\% improvement on smaller networks ($[0, 100)$ size class);
    \item In the IT dataset our multi-layer methodology outperforms the baseline in all size classes, with the maximum performance gain (20\%) on medium networks ($[100, 1000)$ size class); the baseline generally reaches bad performances compared to the US scenario.
\end{itemize}
Overall, our performances are comparable with those achieved by two state-of-the-art deep learning models for "fake news" detection \cite{monti2019fake}\cite{Grover}.

\subsection{Layer importance analysis}
In order to understand the impact of each layer on the performances of classifiers, we performed additional experiments considering separately each layer (we ignored \textbf{T} and \textbf{U} features relative to pure tweets). 

In Table 5 we show metrics for each layer and all size classes, computed with a 10-fold stratified shuffle split cross validation, evaluated on the US dataset; in Figure 5 we show AUROC values for each layer compared with the general multi-layer approach. We can notice that both Q and M layers alone capture adequately the discrepancies of the two distinct news domains in the United States as they obtain good results with AUROC values in the range 75\%-86\%; these are comparable with those of the multi-layer approach which, nevertheless, outperforms them across all size classes.

We obtained similar performances for the Italian dataset, as the M layer obtains comparable performances w.r.t multi-layer approach with AUROC values in the range 72\%-82\%. We do not show these results for sake of conciseness.

\begin{table}[!t]
\centering
\resizebox{\linewidth}{!}{%
\begin{tabular}{l|l|l|l|l|l|}
\textbf{Size Class} & \textbf{Metric}  & \textbf{Quotes} & \textbf{Retweets} & \textbf{Mentions} & \textbf{Replies} \\
\hline
\multirow{3}{*}{$[0, 100$)} &
AUROC
 & \textbf{0.75 $\pm$ 0.02}
 & 0.63 $\pm$ 0.02
 & \textbf{0.75 $\pm$ 0.02}
 & 0.61 $\pm$ 0.02
\\ \cline{2-6} &
Precision
 & \textbf{0.71 $\pm$ 0.02}
 & 0.59 $\pm$ 0.02
 & 0.70 $\pm$ 0.02
 & 0.60 $\pm$ 0.04
\\ \cline{2-6} &
Recall
 & 0.66 $\pm$ 0.01
 & 0.55 $\pm$ 0.01
 & \textbf{0.67 $\pm$ 0.01}
 & 0.54 $\pm$ 0.02
\\ \cline{2-6} &
F1-score
 & 0.66 $\pm$ 0.02
 & 0.53 $\pm$ 0.02
 & \textbf{0.68 $\pm$ 0.02}
 & 0.50 $\pm$ 0.06
\\ \hline
\multirow{3}{*}{$[100, 1000)$} &
AUROC
 & \textbf{0.81 $\pm$ 0.02}
 & 0.63 $\pm$ 0.02
 & \textbf{0.81 $\pm$ 0.02}
 & 0.65 $\pm$ 0.03
\\ \cline{2-6} &
Precision
 & 0.73 $\pm$ 0.02
 & 0.61 $\pm$ 0.02
 & \textbf{0.75 $\pm$ 0.02}
 & 0.65 $\pm$ 0.02
\\ \cline{2-6} &
Recall
 & 0.73 $\pm$ 0.02
 & 0.60 $\pm$ 0.02
 & \textbf{0.75 $\pm$ 0.02}
 & 0.62 $\pm$ 0.02
\\ \cline{2-6} &
F1-score
 & 0.73 $\pm$ 0.02
 & 0.60 $\pm$ 0.02
 & \textbf{0.75 $\pm$ 0.02}
 & 0.60 $\pm$ 0.02
\\ \hline
\multirow{3}{*}{$[1000, +\infty)$} &
AUROC
 & \textbf{0.85 $\pm$ 0.08}
 & 0.62 $\pm$ 0.08
 & 0.84 $\pm$ 0.04
 & 0.66 $\pm$ 0.06
\\ \cline{2-6} &
Precision
 & \textbf{0.80 $\pm$ 0.08}
 & 0.61 $\pm$ 0.08
 & 0.75 $\pm$ 0.06
 & 0.61 $\pm$ 0.10
\\ \cline{2-6} &
Recall
 & \textbf{0.80 $\pm$ 0.08}
 & 0.60 $\pm$ 0.07
 & 0.75 $\pm$ 0.06
 & 0.59 $\pm$ 0.07
\\ \cline{2-6} &
F1-score
 & \textbf{0.79 $\pm$ 0.08}
 & 0.59 $\pm$ 0.08
 & 0.75 $\pm$ 0.06
 & 0.58 $\pm$ 0.09
\\ \hline
\multirow{3}{*}{$[0, +\infty)$} &
AUROC
 & 0.76 $\pm$ 0.01
 & 0.62 $\pm$ 0.01
 & \textbf{0.77 $\pm$ 0.01}
 & 0.59 $\pm$ 0.04
\\ \cline{2-6} &
Precision
 & 0.70 $\pm$ 0.01
 & 0.58 $\pm$ 0.01
 & \textbf{0.73 $\pm$ 0.01}
 & 0.59 $\pm$ 0.05
\\ \cline{2-6} &
Recall
 & 0.69 $\pm$ 0.01
 & 0.56 $\pm$ 0.01
 & \textbf{0.71 $\pm$ 0.01}
 & 0.55 $\pm$ 0.03
\\ \cline{2-6} &
F1-score
 & 0.69 $\pm$ 0.01
 & 0.53 $\pm$ 0.01
 & \textbf{0.71 $\pm$ 0.01}
 & 0.52 $\pm$ 0.05
\\ \hline
\end{tabular}
}
%
\caption{Different evaluations metrics for LR classifier evaluated on different size classes of the US dataset and trained using features separately for each layer. Best scores for each row are written in bold.}
\end{table}

\begin{figure}[!t]
    \centering
    \includegraphics[width=\linewidth]{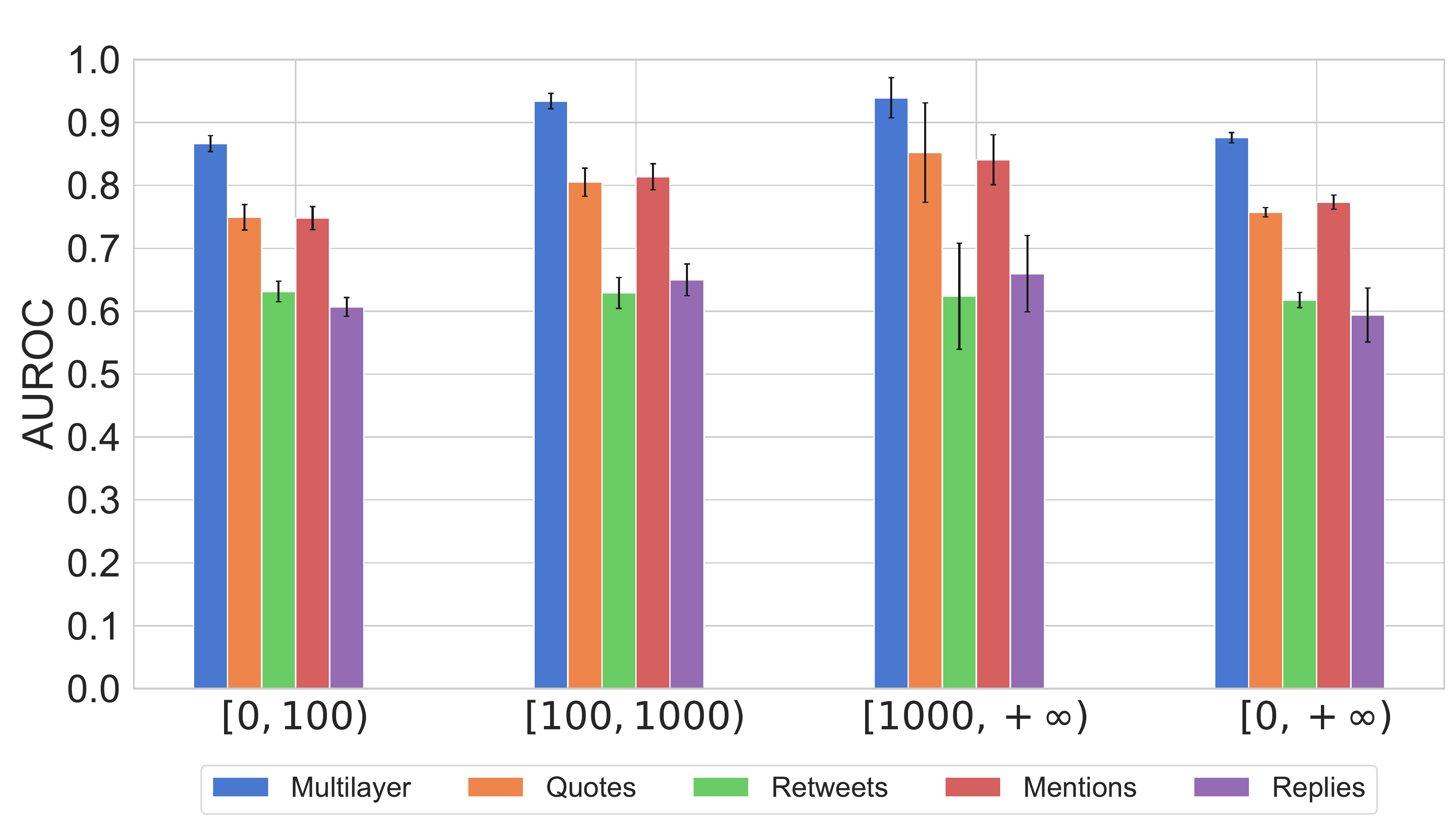}
    \caption{AUROC values for the LR classifier (evaluated on different size classes of the US dataset) trained using different layers separately and aggregated (our multi-layer approach). Error bars indicate the standard deviation of AUROC values over different folds of the cross validation.}
\end{figure}

\subsection{Feature importance analysis}
We further investigated the importance of each feature by performing a $\chi^2$ test, with 10-fold stratified shuffle split cross validation, considering the entire range of network sizes $[0, +\infty)$. We show the Top-5 most discriminative features for each country in Table 6.

We can notice the exact same set of features (with different relative orderings in the Top-3) in both countries; these correspond to two global network propertie--LWCC, which indicates the size of the largest cascade in the layer, and SCC, which correlates with the size of the network--associated to the same set of layers (Quotes, Retweets and Mentions).

We further performed a $\chi^2$ test to highlight the most discriminative features in the M layer of both countries, which performed equally well in the classification task as previously highlighted; also in this case we focused on the entire range of network sizes $[0, +\infty)$. Interestingly, we discovered exactly the same set of Top-3 features in both countries, namely LWCC, SCC and DWCC (which indicates the depth of the largest cascade in the layer).

An inspection of the distributions of all aforementioned features revealed that disinformation news exhibit \textit{on average} larger values than mainstream news\footnote{We also performed a Kolmogorov-Smirnov two-sample test to assess whether distributions of these features are statistically equivalent across the two news domains; the hypothesis was rejected in all cases at $\alpha=0.05$.}.

We can qualitatively sum up these results as follows:
\begin{enumerate}
    \item Sharing patterns in the two news domains exhibit discrepancies which might be country-independent and due to the content that is being shared.
    \item Interactions in disinformation sharing cascades tends to be broader and deeper than in mainstream news, as widely reported in the literature \cite{Vosoughi18}\cite{Bovet2019}\cite{spreading2016}.
    \item Users likely make a different usage of mentions when sharing news belonging to the two domains, consequently shaping different sharing patterns.
\end{enumerate}

\begin{table}[!t]
\centering
\resizebox{\linewidth}{!}{%
\begin{tabular}{lll}
Rank & {\textbf{US}} & {\textbf{IT}} \\
 \hline
\#1 & SCC (Quotes) & LWCC (Retweets) \\
\#2 & LWCC (Retweets) & SCC (Retweets)\\
\#3 & SCC (Retweets) & SCC (Quotes)\\
\#4 & LWCC (Quotes) & LWCC (Quotes)\\
\#5 & LWCC (Mentions) & LWCC (Mentions)\\
\hline
\end{tabular}%
}
\caption{Top-5 most discriminative features according to $\chi^2$ test evaluated on both US and IT datasets (considering networks in the $[0, +\infty)$ size class).}
\end{table}

\begin{figure}[!t]
    \centering
    \includegraphics[width=\linewidth]{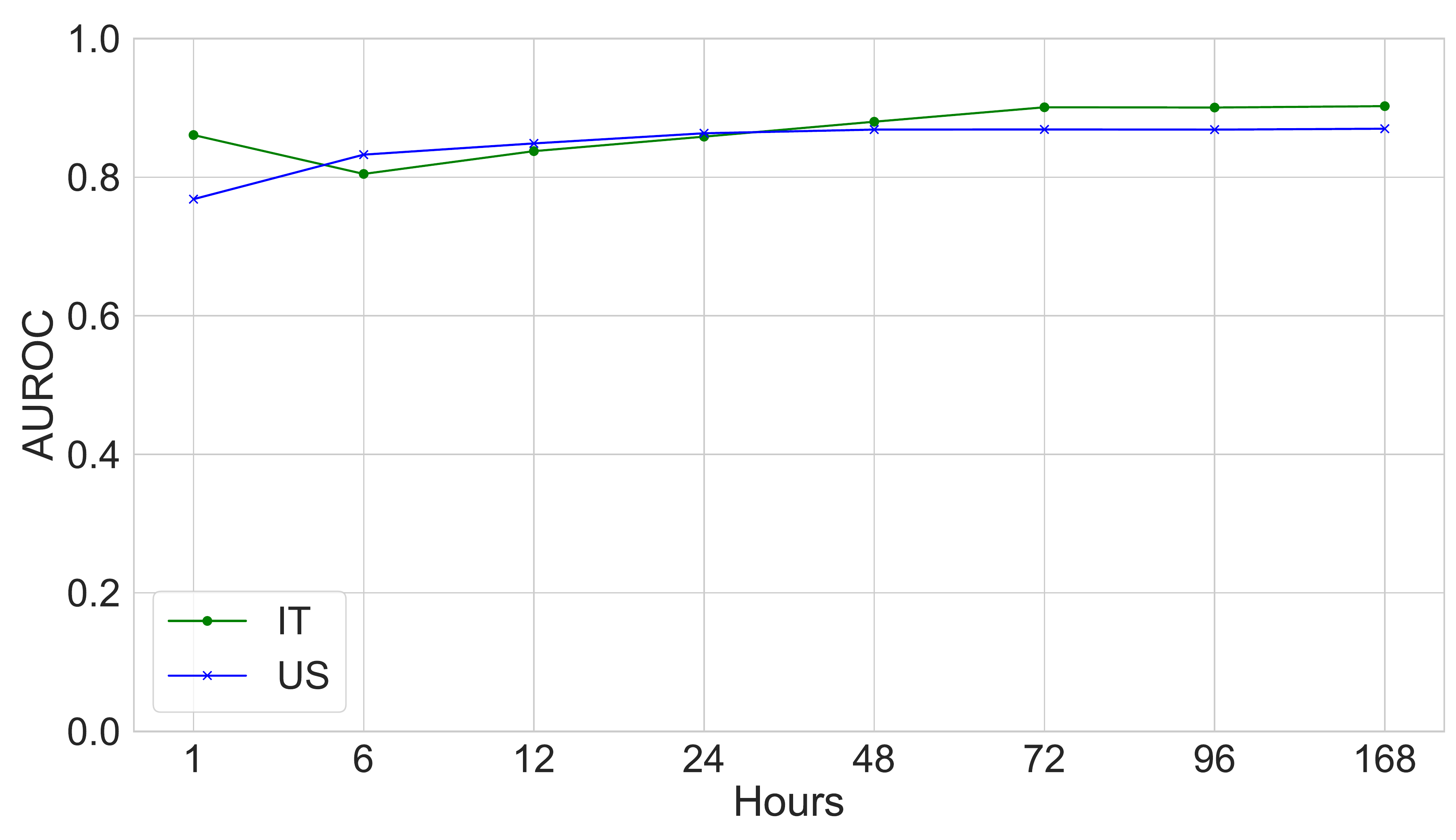}
    \caption{AUROC values for LR classifier evaluated on US (blue) and IT (green) dataset, considering different lifetimes (or spreading duration) for multi-layer networks. Error bars indicate standard deviations of 10-fold cross validation.}
\end{figure}
\subsection{Temporal analysis}
Similar to \cite{monti2019fake}, we carried out additional experiments to answer the following question: how long do we need to observe a news spreading on Twitter in order to accurately classify it as disinformation or mainstream?

With this goal, we built several versions of our original dataset of multi-layer networks by considering in turn the following lifetimes\footnote{For each news article we built the corresponding multi-layer network considering only tweets shared in the 1st hour, the first 6 hours, the first 12 hours, etc.}: 1 hour, 6 hours, 12 hours, 1 day, 2 days, 3 days and 7 days; for each case, we computed the global network properties of the corresponding network and evaluated the LR classifier with 10-fold cross validation, separately for each lifetime (and considering always the entire set of networks). We show corresponding AUROC values for both US and IT datasets in Figure 6.

We can see that in both countries news diffusion networks can be accurately classified after just a few hours of spreading, with AUROC values which are larger than 80\% after only 6 hours of diffusion. These results are very promising and suggest that articles pertaining to the two news domains exhibit discrepancies in their sharing patterns that can be timely exploited in order to rapidly detect misleading items from factual information.

\section{Conclusions}
In this work we tackled the problem of the automatic classification of news articles in two domains, namely \textit{mainstream} and \textit{disinformation} news, with a language-independent approach which is based solely on the diffusion of news items on Twitter social platform. We disentangled different types of interactions on Twitter to accordingly build a multi-layer representation of news diffusion networks, and we computed a set of global network properties--separately for each layer--in order to encode each network with a tuple of features. Our goal was to investigate whether a multi-layer representation performs better than one layer \cite{PierriScirep2019}, and to understand which of the features, observed at given layers, are most effective in the classification task.

Experiments with an off-the-shelf classifier such as Logistic Regression on datasets pertaining to two different media landscapes (US and Italy) yield very accurate classification results (AUROC up to 94\%), even when accounting for the different political bias of news sources, which are far better than our baseline \cite{PierriScirep2019} with improvements up to 20\%.
Classification performances using single layers show that the layer of mentions alone entails better performance
w.r.t other layers in both countries.

We also highlighted the most discriminative features across different layers in both countries; the results suggest that differences between the two news domains might be country-independent but rather due only to the typology of content shared, and that disinformation news shape broader and deeper cascades.

Additional experiments involving the temporal evolution of Twitter diffusion networks show that our methodology can accurate classify mainstream and disinformation news after a few hours of propagation on the platform.

Overall, our results prove that the topological features of multi-layer diffusion networks might be effectively exploited to detect online disinformation. We do not deny the presence of deceptive efforts to orchestrate the regular spread of information on social media via content amplification and manipulation \cite{starbird}\cite{badawy2018}. On the contrary, we postulate that such hidden forces might play to accentuate the discrepancies between the diffusion patterns of disinformation and mainstream news (and thus to make our methodology effective).


In the future we aim to further investigate three directions: (1) employ temporal networks to represent news diffusion and apply classification techniques that take into account the sequential aspect of data (e.g. recurrent neural networks); (2) carry out an extensive comparison of the diffusion of disinformation and mainstream news across countries to investigate deeper the presence of differences and similarities in sharing patterns; (3) leverage our network-based features in addition to state-of-the-art text-based approaches for "fake news" dete ction in order to deliver a real-world system to detect misleading and harmful information spreading on social media.

\bibliography{bib.bib}

\begin{thebibliography}{}

\bibitem[\protect\citeauthoryear{Allcott and Gentzkow}{2017}]{allcott2017}
Allcott, H., and Gentzkow, M.
\newblock 2017.
\newblock Social media and fake news in the 2016 election.
\newblock {\em Journal of Economic Perspectives} 31(2):211--36.

\bibitem[\protect\citeauthoryear{Badawy, Ferrara, and
  Lerman}{2018}]{badawy2018}
Badawy, A.; Ferrara, E.; and Lerman, K.
\newblock 2018.
\newblock Analyzing the digital traces of political manipulation: the 2016
  russian interference twitter campaign.
\newblock In {\em 2018 IEEE/ACM International Conference on Advances in Social
  Networks Analysis and Mining (ASONAM)},  258--265.
\newblock IEEE.

\bibitem[\protect\citeauthoryear{Barber{\'a} \bgroup et al\mbox.\egroup
  }{2015}]{barbera}
Barber{\'a}, P.; Jost, J.~T.; Nagler, J.; Tucker, J.~A.; and Bonneau, R.
\newblock 2015.
\newblock Tweeting from left to right: Is online political communication more
  than an echo chamber?
\newblock {\em Psychological science} 26(10):1531--1542.

\bibitem[\protect\citeauthoryear{Batagelj and Zaversnik}{2003}]{k-core}
Batagelj, V., and Zaversnik, M.
\newblock 2003.
\newblock An o(m) algorithm for cores decomposition of networks.
\newblock {\em arXiv preprint cs/0310049}.

\bibitem[\protect\citeauthoryear{Bovet and Makse}{2019}]{Bovet2019}
Bovet, A., and Makse, H.~A.
\newblock 2019.
\newblock {Influence of fake news in Twitter during the 2016 US presidential
  election}.
\newblock {\em Nature Communications} 10(1):7.

\bibitem[\protect\citeauthoryear{Bovet, Morone, and Makse}{2018}]{bovet2018bis}
Bovet, A.; Morone, F.; and Makse, H.~A.
\newblock 2018.
\newblock Validation of twitter opinion trends with national polling
  aggregates: Hillary clinton vs donald trump.
\newblock {\em Scientific reports} 8(1):8673.

\bibitem[\protect\citeauthoryear{Boyd, Golder, and Lotan}{2010}]{boyd2010}
Boyd, D.; Golder, S.; and Lotan, G.
\newblock 2010.
\newblock Tweet, tweet, retweet: Conversational aspects of retweeting on
  twitter.
\newblock In {\em 2010 43rd Hawaii International Conference on System
  Sciences},  1--10.
\newblock IEEE.

\bibitem[\protect\citeauthoryear{Conover \bgroup et al\mbox.\egroup
  }{2012}]{conover2012}
Conover, M.~D.; Gon{\c{c}}alves, B.; Flammini, A.; and Menczer, F.
\newblock 2012.
\newblock Partisan asymmetries in online political activity.
\newblock {\em EPJ Data Science} 1(1):6.

\bibitem[\protect\citeauthoryear{Davis \bgroup et al\mbox.\egroup
  }{2016}]{botometer}
Davis, C.~A.; Varol, O.; Ferrara, E.; Flammini, A.; and Menczer, F.
\newblock 2016.
\newblock Botornot: A system to evaluate social bots.
\newblock In {\em Proceedings of the 25th International Conference Companion on
  World Wide Web},  273--274.
\newblock International World Wide Web Conferences Steering Committee.

\bibitem[\protect\citeauthoryear{Del~Vicario \bgroup et al\mbox.\egroup
  }{2016}]{spreading2016}
Del~Vicario, M.; Bessi, A.; Zollo, F.; Petroni, F.; Scala, A.; Caldarelli, G.;
  Stanley, H.~E.; and Quattrociocchi, W.
\newblock 2016.
\newblock The spreading of misinformation online.
\newblock {\em Proceedings of the National Academy of Sciences}
  113(3):554--559.

\bibitem[\protect\citeauthoryear{Fawcett}{2006}]{roc}
Fawcett, T.
\newblock 2006.
\newblock An introduction to roc analysis.
\newblock {\em Pattern recognition letters} 27(8):861--874.

\bibitem[\protect\citeauthoryear{Goel \bgroup et al\mbox.\egroup }{2015}]{goel}
Goel, S.; Anderson, A.; Hofman, J.; and Watts, D.~J.
\newblock 2015.
\newblock The structural virality of online diffusion.
\newblock {\em Management Science} 62(1):180--196.

\bibitem[\protect\citeauthoryear{Grabowicz \bgroup et al\mbox.\egroup
  }{2012}]{mentions}
Grabowicz, P.~A.; Ramasco, J.~J.; Moro, E.; Pujol, J.~M.; and Eguiluz, V.~M.
\newblock 2012.
\newblock Social features of online networks: The strength of intermediary ties
  in online social media.
\newblock {\em PloS one} 7(1):e29358.

\bibitem[\protect\citeauthoryear{Grinberg \bgroup et al\mbox.\egroup
  }{2019}]{Grinberg}
Grinberg, N.; Joseph, K.; Friedland, L.; Swire-Thompson, B.; and Lazer, D.
\newblock 2019.
\newblock Fake news on twitter during the 2016 u.s. presidential election.
\newblock {\em Science} 363(6425):374--378.

\bibitem[\protect\citeauthoryear{Hagberg, Swart, and S~Chult}{2008}]{networkx}
Hagberg, A.; Swart, P.; and S~Chult, D.
\newblock 2008.
\newblock Exploring network structure, dynamics, and function using networkx.
\newblock Technical report, Los Alamos National Lab.(LANL), Los Alamos, NM
  (United States).

\bibitem[\protect\citeauthoryear{Kivel{\"a} \bgroup et al\mbox.\egroup
  }{2014}]{multilayer}
Kivel{\"a}, M.; Arenas, A.; Barthelemy, M.; Gleeson, J.~P.; Moreno, Y.; and
  Porter, M.~A.
\newblock 2014.
\newblock Multilayer networks.
\newblock {\em Journal of complex networks} 2(3):203--271.

\bibitem[\protect\citeauthoryear{Lazer \bgroup et al\mbox.\egroup
  }{2018}]{Lazer18}
Lazer, D. M.~J.; Baum, M.~A.; Benkler, Y.; Berinsky, A.~J.; Greenhill, K.~M.;
  Menczer, F.; Metzger, M.~J.; Nyhan, B.; Pennycook, G.; Rothschild, D.;
  Schudson, M.; Sloman, S.~A.; Sunstein, C.~R.; Thorson, E.~A.; Watts, D.~J.;
  and Zittrain, J.~L.
\newblock 2018.
\newblock The science of fake news.
\newblock {\em Science} 359(6380):1094--1096.

\bibitem[\protect\citeauthoryear{Lema{{\^i}}tre, Nogueira, and
  Aridas}{2017}]{imbalanced}
Lema{{\^i}}tre, G.; Nogueira, F.; and Aridas, C.~K.
\newblock 2017.
\newblock Imbalanced-learn: A python toolbox to tackle the curse of imbalanced
  datasets in machine learning.
\newblock {\em Journal of Machine Learning Research} 18(17):1--5.

\bibitem[\protect\citeauthoryear{Mitchell \bgroup et al\mbox.\egroup
  }{2014}]{pew}
Mitchell, A.; Gottfried, J.; Kiley, J.; and Matsa, K.~E.
\newblock 2014.
\newblock Political polarization \& media habits.
\newblock {\em Pew Research Center} 21.

\bibitem[\protect\citeauthoryear{Monti \bgroup et al\mbox.\egroup
  }{2019}]{monti2019fake}
Monti, F.; Frasca, F.; Eynard, D.; Mannion, D.; and Bronstein, M.~M.
\newblock 2019.
\newblock Fake news detection on social media using geometric deep learning.
\newblock {\em arXiv preprint arXiv:1902.06673}.

\bibitem[\protect\citeauthoryear{Morstatter \bgroup et al\mbox.\egroup
  }{2013}]{twittersample}
Morstatter, F.; Pfeffer, J.; Liu, H.; and Carley, K.~M.
\newblock 2013.
\newblock Is the sample good enough? comparing data from twitter's streaming
  api with twitter's firehose.
\newblock In {\em Seventh international AAAI conference on weblogs and social
  media}.

\bibitem[\protect\citeauthoryear{Nickerson}{1998}]{confirmationbias}
Nickerson, R.~S.
\newblock 1998.
\newblock Confirmation bias: A ubiquitous phenomenon in many guises.
\newblock {\em Review of General Psychology} 2(2):175.

\bibitem[\protect\citeauthoryear{Nielsen \bgroup et al\mbox.\egroup
  }{2019}]{reuters2019}
Nielsen, R.~K.; Newman, N.; Fletcher, R.; and Kalogeropoulos, A.
\newblock 2019.
\newblock Reuters institute digital news report 2019.
\newblock {\em Report of the Reuters Institute for the Study of Journalism}.

\bibitem[\protect\citeauthoryear{Pedregosa \bgroup et al\mbox.\egroup
  }{2011}]{scikit}
Pedregosa, F.; Varoquaux, G.; Gramfort, A.; Michel, V.; Thirion, B.; Grisel,
  O.; Blondel, M.; Prettenhofer, P.; Weiss, R.; Dubourg, V.; et~al.
\newblock 2011.
\newblock Scikit-learn: Machine learning in python.
\newblock {\em Journal of machine learning research} 12(Oct):2825--2830.

\bibitem[\protect\citeauthoryear{Pierri, Artoni, and
  Ceri}{2020}]{PierriArtoni2019}
Pierri, F.; Artoni, A.; and Ceri, S.
\newblock 2020.
\newblock Investigating italian disinformation spreading on twitter in the
  context of 2019 european elections.
\newblock {\em PLoS One} 15(1):e0227821, 2020.

\bibitem[\protect\citeauthoryear{Pierri, Piccardi, and
  Ceri}{2020}]{PierriScirep2019}
Pierri, F.; Piccardi, C.; and Ceri, S.
\newblock 2020.
\newblock Topology comparison of twitter diffusion networks effectively reveals
  misleading news.
\newblock {\em Scientific Reports} 10:1372, 2020 .

\bibitem[\protect\citeauthoryear{Ratkiewicz \bgroup et al\mbox.\egroup
  }{2011}]{astroturfing}
Ratkiewicz, J.; Conover, M.; Meiss, M.; Gon{\c{c}}alves, B.; Patil, S.;
  Flammini, A.; and Menczer, F.
\newblock 2011.
\newblock {Detecting and Tracking Political Abuse in Social Media}.
\newblock {\em ICWSM 2011}  249.

\bibitem[\protect\citeauthoryear{Reed, Turiel, and Brown}{2013}]{naiverealism}
Reed, E.~S.; Turiel, E.; and Brown, T.
\newblock 2013.
\newblock Naive realism in everyday life: Implications for social conflict and
  misunderstanding.
\newblock In {\em Values and knowledge}. Psychology Press.
\newblock  113--146.

\bibitem[\protect\citeauthoryear{Saram{\"a}ki \bgroup et al\mbox.\egroup
  }{2007}]{cc}
Saram{\"a}ki, J.; Kivel{\"a}, M.; Onnela, J.-P.; Kaski, K.; and Kertesz, J.
\newblock 2007.
\newblock Generalizations of the clustering coefficient to weighted complex
  networks.
\newblock {\em Physical Review E} 75(2):027105.

\bibitem[\protect\citeauthoryear{Shao \bgroup et al\mbox.\egroup
  }{2016}]{hoaxy}
Shao, C.; Ciampaglia, G.~L.; Flammini, A.; and Menczer, F.
\newblock 2016.
\newblock Hoaxy: A platform for tracking online misinformation.
\newblock In {\em Proceedings of the 25th International Conference Companion on
  World Wide Web}, WWW '16 Companion,  745--750.
\newblock Republic and Canton of Geneva, Switzerland: International World Wide
  Web Conferences Steering Committee.

\bibitem[\protect\citeauthoryear{Shao \bgroup et al\mbox.\egroup
  }{2018a}]{botnature}
Shao, C.; Ciampaglia, G.~L.; Varol, O.; Yang, K.-C.; Flammini, A.; and Menczer,
  F.
\newblock 2018a.
\newblock The spread of low-credibility content by social bots.
\newblock {\em Nature Communications} 9(1):4787.

\bibitem[\protect\citeauthoryear{Shao \bgroup et al\mbox.\egroup
  }{2018b}]{misinformation}
Shao, C.; Hui, P.-M.; Wang, L.; Jiang, X.; Flammini, A.; Menczer, F.; and
  Ciampaglia, G.~L.
\newblock 2018b.
\newblock Anatomy of an online misinformation network.
\newblock {\em PLOS ONE} 13(4):1--23.

\bibitem[\protect\citeauthoryear{Stewart, Arif, and Starbird}{2018}]{starbird}
Stewart, L.~G.; Arif, A.; and Starbird, K.
\newblock 2018.
\newblock Examining trolls and polarization with a retweet network.
\newblock In {\em Proceedings ACM WSDM, Workshop on Misinformation and
  Misbehavior Mining on the Web}.

\bibitem[\protect\citeauthoryear{Vicario \bgroup et al\mbox.\egroup
  }{2019}]{vicario2017}
Vicario, M.~D.; Quattrociocchi, W.; Scala, A.; and Zollo, F.
\newblock 2019.
\newblock Polarization and fake news: Early warning of potential misinformation
  targets.
\newblock {\em ACM Transactions on the Web (TWEB)} 13(2):10.

\bibitem[\protect\citeauthoryear{Vosoughi, Roy, and Aral}{2018}]{Vosoughi18}
Vosoughi, S.; Roy, D.; and Aral, S.
\newblock 2018.
\newblock The spread of true and false news online.
\newblock {\em Science} 359(6380):1146--1151.

\bibitem[\protect\citeauthoryear{Wasserman, Faust, and
  others}{1994}]{sociologists}
Wasserman, S.; Faust, K.; et~al.
\newblock 1994.
\newblock {\em Social network analysis: Methods and applications}, volume~8.
\newblock Cambridge university press.

\bibitem[\protect\citeauthoryear{Wiener}{1947}]{wiener}
Wiener, H.
\newblock 1947.
\newblock Structural determination of paraffin boiling points.
\newblock {\em Journal of the American Chemical Society} 69(1):17--20.

\bibitem[\protect\citeauthoryear{Zellers \bgroup et al\mbox.\egroup
  }{2019}]{Grover}
Zellers, R.; Holtzman, A.; Rashkin, H.; Bisk, Y.; Farhadi, A.; Roesner, F.; and
  Choi, Y.
\newblock 2019.
\newblock Defending against neural fake news.
\newblock {\em arXiv preprint arXiv:1905.12616}.

\bibitem[\protect\citeauthoryear{Zhao \bgroup et al\mbox.\egroup
  }{2018}]{chinafake}
Zhao, Z.; Zhao, J.; Sano, Y.; Levy, O.; Takayasu, H.; Takayasu, M.; Li, D.; and
  Havlin, S.
\newblock 2018.
\newblock Fake news propagate differently from real news even at early stages
  of spreading.
\newblock {\em arXiv preprint arXiv:1803.03443}.

\end{thebibliography}
\bibliographystyle{aaai}

\end{document}